\documentstyle{article}

\def\mup{m_{\scriptscriptstyle u}}
\def\mc{m_{\scriptscriptstyle c}}
\def\mt{m_{\scriptscriptstyle t}}
\def\md{m_{\scriptscriptstyle d}}

\def\mb{m_{\scriptscriptstyle b}}
\def\mtau{m_{\scriptscriptstyle \tau}}
\def\MG{M_{\scriptscriptstyle G}}
\def\mZ{  m_{ {\scriptscriptstyle Z} }     }

\def\alphas{\alpha_{\scriptscriptstyle s}}

\def\Vckm{V_{\scriptscriptstyle CKM} }
\def\Vcb{V_{\scriptscriptstyle cb} }
\def\Vub{V_{\scriptscriptstyle ub} }

\def\VLu{V_{\scriptscriptstyle u}^{\scriptscriptstyle L} }
\def\VRu{V_{\scriptscriptstyle u}^{\scriptscriptstyle R} }
\def\VLd{V_{\scriptscriptstyle d}^{\scriptscriptstyle L} }
\def\VRd{V_{\scriptscriptstyle d}^{\scriptscriptstyle R} }
\def\Ve{V_{\scriptscriptstyle e}}
\def\gc{g_{\scriptscriptstyle c} }
\def\Tr{\mathop{\rm Tr}}

\font\newsymb=msxm10   \def\lapprox{\mathop{\hbox{\newsymb .}}}
                       
\begin{document}
\begin{center}
\today      \hfill       OHSTPY-HEP-92-018\\
               \hfill       LBL-32817\\
               \hfill       UCB-PTH-92/28\\
              \hfill       Stanford-92-25\\
{\bf \large  Precise Predictions for $\mt$, $\Vcb$, and $\tan\beta$ }
\footnote{This work is supported by NSF Grants PHY-89-17438
and PHY-90-21139  at Stanford and Berkeley and by DOE-AC03-76SF00098
and DOE-ER-01545-585 at Berkeley and Ohio State.}\\
\vskip 1.0 cm

       {
      {\bf Greg W. Anderson and Stuart Raby\footnote{On leave of
absence from Theoretical Division, Los Alamos National Laboratory,
Los Alamos  NM 87545} }
           \vskip 0.25 cm
     {\it Department of Physics,
          The Ohio State University\\
           Columbus, OH 43210.    }
           \vskip 0.5cm
      {\bf Savas Dimopoulos}
           \vskip 0.25 cm
     {\it Department of Physics,
          Stanford University,
          Stanford, CA 94305. }\\
     {\bf Lawrence J. Hall}
           \vskip 0.25 cm
     {\it  Physics Department \& LBL,
          University of California, Berkeley\\
         Berkeley, CA 94720  } }

\end{center}
\newpage
\setcounter{page}{1}

\begin{abstract}
The fermion mass and mixing angle predictions of
a recently proposed framework
are investigated for large b and $\tau$ Yukawa couplings.
A new allowed region of parameters is found for this large $\tan \beta$
case.
The two predictions which are substantially altered,
$m_t$ and $\tan \beta$, are displayed, including the dependence on the inputs
$|V_{cb}|$, $m_c$, $m_b$ and $\alpha_s$.
A simple restriction on this framework yields an additional prediction, for
$|V_{cb}|$.  If the  b,t, and $\tau$ Yukawas are equal at the GUT scale
then $|\Vcb|$ is predicted and the top quark mass is constrained to lie
in the range $\mt = 179 \pm 4$ GeV.

\end{abstract}
\newpage

\section{Introduction}
\def\theequation{1.\arabic{equation}}
\indent

The majority of the parameters of the standard model, thirteen out of eighteen,
originate in the mass matrices of the quarks and charged leptons.
The most promising idea which has led to predictions of some of these flavor
parameters is the reduction of parameters made possible by a combination of
grand unified symmetries \cite{gg} and flavor symmetries \cite{wf}. The most
successful prediction of any parameter of the standard model is that of the
weak mixing angle in supersymmetric grand unified theories (GUTs) \cite{gqw}:
the predicted value of .233$\pm$.002 should be compared  with the experimental
value of .233$\pm$.001. The first flavor parameter to be predicted in GUTs was
$m_b/m_\tau$ \cite{ceg}. This prediction is affected by the mass of the heavy
top quark and again favors supersymmetric unification.\cite{ILM}

Recently three of us introduced the most predictive scheme yet constructed for
flavor parameters \cite{dhr}, based on the family symmetries of Georgi and
Jarlskog \cite{gj} and on supersymmetric GUTs. Including $\tan \beta$, the
ratio of electroweak vevs, there are fourteen flavor parameters, and these are
given in terms of only eight GUT parameters, yielding six predictions. To
obtain these predictions the Yukawa matrices must be scaled with the
renormalization group (RG) from GUT to weak scales. Previously this was done
by including the one-loop RG scaling effects induced by gauge and top quark
Yukawa interactions \cite{dhr,bbhz}. This is sufficient only for moderate
values of $\tan \beta$, since as $\tan \beta$ is increased, the b and $\tau$
Yukawa couplings increase and contribute to the RG scaling.

In this paper we reanalyze the DHR framework\cite{dhr}
 using the full one-loop RG
equations, thus including RG scaling effects induced by the b and $\tau$ Yukawa
couplings. There are several reasons for doing this. Previous analyses
\cite{dhr,bbhz}
 showed that $\tan \beta$ cannot be less than about two
when the strong coupling constant is $\alphas(\mZ)=.109$.
Furthermore as $\tan \beta$ is increased so $V_{cb}$ is decreased, improving
the agreement with experiment. It is clearly of great interest to know whether
this agreement can be further improved by going to values of $\tan \beta$
outside the range of validity of previous calculations. Furthermore the
previous predictions for $m_t$ will be affected by large b and $\tau$ Yukawa
couplings, which contribute to the RG equation for the t Yukawa coupling.
Finally, studying the case of large b and $\tau$ Yukawa couplings allows the
exploration of a particularly simple restriction on the DHR Yukawa
couplings; namely that the GUT scale b and t Yukawa couplings are equal.

\section{Evolving the DHR Texture}
\def\theequation{2.\arabic{equation}}
\setcounter{equation}{0}
\indent

The DHR ans\"atze for the three Yukawa matrices at the
renormalization group scale $\mu = \MG$ is
 \begin{eqnarray}
{\bf U} =  \pmatrix{  0  & C  & 0 \cr  C &  0  & B  \cr  0  & B &  A \cr},
{\bf D} =  \pmatrix{  0  & Fe^{i\phi} &  0 \cr  Fe^{-i\phi} & E & 0 \cr 0
    & 0 & D \cr},
{\bf E} =  \pmatrix{  0  & F &  0 \cr  F & -3 E & 0 \cr 0 &
0 & D \cr} .\nonumber
\end{eqnarray}
Below the GUT scale, the symmetries which maintained zero entries
and preserved the relationship between ${\bf D}$ and ${\bf E}$ are
broken. The evolution of the Yukawa matrices between the GUT scale
and the weak scale is given by\cite{Inoue}:

\begin{eqnarray}
\mu \frac{d}{d\mu} {\bf U} &=&
\frac{1}{(4\pi)^2}\left\{ \left[3\Tr \left( {\bf U}{\bf U}^{\dag}
\right) -\left(\frac{16}{3}\gc^2+3g^2 +
\frac{13}{9}g'^2\right)\right] {\bf U}\right. \nonumber \\  &+&
\left. 3{\bf U}{\bf U}^{\dag}{\bf U} + {\bf D}{\bf D}^{\dag}{\bf
U}\right\} \nonumber
 \hskip 6cm (2.2a) \\
   \mu \frac{d}{d\mu} {\bf D} &=&
\frac{1}{(4\pi)^2}\left\{ \left[\Tr \left( {\bf E}{\bf E}^{\dag}+
3{\bf D}{\bf D}^{\dag}\right) -\left(\frac{16}{3}\gc^2+3g^2 +
\frac{7}{9}g'^2\right)\right]{\bf D} \right. \nonumber \\
&+& \left. 3{\bf D}{\bf D}^{\dag}{\bf D}+
{\bf U}{\bf U}^{\dag}{\bf D}\right\} \nonumber \hskip 6cm (2.2b) \\
    \mu \frac{d}{d\mu} {\bf E} &=&
\frac{1}{(4\pi)^2}\left\{ \left[\Tr \left( {\bf E}{\bf E}^{\dag}+
3{\bf D}{\bf D}^{\dag}\right) -\left(3g^2 + 3g'^2\right)\right]
{\bf E}\right. \nonumber \\
&+& \left. 3{\bf E}{\bf E}^{\dag}{\bf E}\right\} .  \nonumber
\hskip 8cm (2.2c)
\end{eqnarray}

In general, ${\bf U}$ and
${\bf D}$
are neither symmetric or hermitian below the GUT scale.  We stop the
RGE evolution at a scale $\mu \simeq \mt$.
The weak scale quark and lepton mass matrices are then given by
\begin{equation}
m_{U} = {\bf U}\frac{v \sin{\beta}}{\sqrt{2} },\,\,\,
m_{D} = {\bf D}\frac{v \cos{\beta}}{\sqrt{2} },\,\,\,
m_{E} = {\bf E}\frac{v \cos{\beta}}{\sqrt{2} },\,\,\,
\end{equation}
where $v = 246$ GeV.
These three weak scale Yukawa matrices are
 diagonalized by the unitary matrices
$\VLu$, $\VRu$, $\VLd$, $\VRd$,
and $\Ve$ defined by ${\bf U}^{diag}=\VLu {\bf U}\VRu^{\dag}$,
${\bf D}^{diag} = \VLd {\bf D} \VRd^{\dag}$ and
${\bf E}^{diag} = \Ve {\bf E} \Ve^{\dag}$.
The Cabibbo-Kobayashi-Maskawa
[CKM] mixing matrix is then given by $\Vckm = \VLu \VLd^{\dag}$.
Below the weak scale we evolve the fermion masses to one-loop
in QED and two-loops in QCD.  We define the resulting mass
enhancement  factors  by $\eta_{f} = m_{f}(m_{f})/m_{f}(\mt)$ for
$f=e,\mu,\tau,c,b$ and $\eta_{f} = m_{f}(1 GeV)/m_{f}(\mt)$ for
$f=u,d,s$.  The dependence of these mass enhancement factors
 on the strong coupling constant, shown in figure 1, has important
consequences.

All of the numerical results quoted in this paper are
a result of running and diagonalizing the Yukawa couplings in
matrix form according to the full one-loop RGE.
The eight parameters $A,B,C,D,E,F,\phi,$ and $\tan{\beta}$ are best
determined from the masses and mixing angles; $m_e,m_\mu,\mtau,\mc
,\mb,\mup/\md,|V_{us}|,$ and $|\Vcb|$.  Consider a fixed but arbitrary
value of $\tan{\beta}$.  Since the RG evolution of the Yukawa matrix
elements can be  significantly affected by all three third generation
Yukawa couplings, the parameters $A$
and $D$ must be determined first.  Both of these couplings can be
reliable determined by the combined constraints of the tau lepton mass and
the $b/\tau$ ratio.  Since the charged lepton
masses are well known and the  RGE for lepton masses is not
complicated by uncertainties in $\alphas$, the couplings $E$ and
$F$ are then very  reliably determined by the electron and muon masses.
The parameter $B$ can then  be fixed by the charm quark mass,  $C$
set  by the up to down ratio, and the phase $\phi$ determined  from
$|V_{us}|$.  Finally two solutions for $\tan{\beta}$ can be found
as a function of $|\Vcb|$.   The smaller branch corresponds to the
solution obtained previously\cite{dhr}, while the larger branch
represents a new solution (see fig. 2).  Once $\tan{\beta}$ is determined
we have a prediction for $\mt$  and the remaining fermion masses and
mixing angles (see fig. 3).  In figure 4, we plot
the prediction for $\mt$ directly as a function of $|\Vcb|$.

How constrained are the predicted values of $\mt$ and $\tan{\beta}$?
Having specified $\mc$, $\mb$ and $\alphas$, choosing a value for
$|\Vcb|$ in principle determines two discrete values for
$\tan{\beta}$.  However,  over a large range, $\tan{\beta}$ is a very
sensitive function of $|\Vcb|$.  In addition, the relationship
between $\tan{\beta}$ and $|\Vcb|$ is affected by uncertainties in
$\alphas , \mb,$ and $\mc$.  Hence, until we know the values
of $|\Vcb| , \mc, \mb$ and $\alphas$ with better precision,
$\tan{\beta}$ is not well determined.  For this reason we chose to
display our predictions for $\tan{\beta}$  and $\mt$ as an allowed
region where the quantities mentioned above are allowed to vary
within the current experimental uncertainties. Moreover, figs.
2(a,b,c), 3(a,b,c) and 4(a,b,c) show the sensitivity of the
predictions to variations in $\alphas(\mZ)$ for the three values,
.110, .118, and .126 .  Note, the plots in this paper are for the
pole mass of the top quark defined in terms of the running mass by
\begin{equation} \mt(pole) = \mt(\mt)\left( 1 +
\frac{4\alphas}{3\pi}\right)  . \end{equation}

Let us now consider the theoretical uncertainty in our predictions
due to the experimental/theoretical uncertainty in $\alphas(\mZ)$.
Although a seemingly  precise value of $\alphas(\mZ)$ can be
obtained by imposing one-loop unification of the three gauge
couplings at a single point $\mu = \MG$ and evolving the strong
coupling down to the weak scale, both GUT and weak scale threshold
corrections introduce significant theoretical uncertainties in
$\alphas$.
We therefore allow $\alpha_s$ to vary over the range
$\alphas = .118 \pm .008$.
The uncertainty in $\alphas$ affects the predictions
through their dependence on the renormalization factors.
For example at fixed $\tan{\beta}$, the square of $|\Vcb|$
is proportional to the bottom quark and charm quark Yukawa
couplings at the scale $\mu = \mt$ (see ref.~\cite{dhr}).  So for
fixed charm and bottom quark masses.
\begin{equation}
|\Vcb| ^2 \propto \frac{1}{\eta_b \eta_c} .
\end{equation}
Thus a larger value of $\alphas$, gives a smaller value of $|\Vcb|$.
Similarly, $|\Vub|/|\Vcb|$ grows like $\sqrt{ \eta_c}$.

The qualitative behavior of the solutions at large values  of
$\tan{\beta}$ can be understood by appealing to the solutions
obtained previously for small values of $\tan{\beta}$ and by
studying the effect of the b-quark Yukawa coupling on the running of
the $b/\tau$ ratio. The renormalization group equation  for the
$\tau$  Yukawa coupling doesn't depend on the top quark Yukawa at
one-loop. So once we specify values of $\tan{\beta}$ and
 $\alphas(\mZ)$ the  GUT scale input $D$ can be given as a unique
function of the parameter $A$ by
demanding that the tau lepton mass is $1.78$ GeV.~\footnote{
Although the RG equation for the tau Yukawa coupling doesn't contain
the top quark Yukawa coupling, $\lambda_{\tau}(\mt)$ will in general
depend on $A$
because the evolution of the b Yukawa coupling, which contributes to
the running of $\lambda_{\tau}$,  depends on $A$.}
The top quark
Yukawa coupling can then be determined from the b-quark
mass.  In general, the top quark Yukawa coupling decreases the
$b/\tau$ ratio, and a phenomenologically acceptable ratio requires
a large top quark Yukawa.

Let us now define R as the ratio of the b quark mass to the tau
lepton mass.  Then
\begin{equation}
\mu \frac{dR}{d \mu} = \frac{R}{16\pi^2}
\left( -d_{i}g_{i}^2 + \lambda_t^2  + 3\lambda_{\tau}^2
\left(R^2 -1\right)\right)
 \end{equation}
where the RG constants $d_i$ are defined in ref.~\cite{dhr}.
Because $R$ is larger than unity for most of the RG evolution,
a large $\tau$ Yukawa coupling tends to
decrease the $b/\tau$ ratio.~\footnote{When $A\lapprox 1.4$,
$R\geq 1$ over the entire range of the RG scale $\mu$.}
So, as $\tan{\beta}$ is increased, because
$D$ must  increase to keep the tau mass constant,
$b/\tau$ decreases.  Since $b/\tau$
decreases, a smaller top quark Yukawa coupling is needed to
maintain a fixed $b/\tau$ ratio.    Thus at large values of
$\tan{\beta}$ the top quark Yukawa coupling, $A$, decreases and so
does $\mt$ (fig. 3).  In addition, as $A$ decreases, $|\Vcb|$
increases,  which results in an upper limit on $\tan\beta$ (fig. 2).
These relationships can also be seen in fig. 4 which displays
the prediction for $\mt$ in terms of $|\Vcb|$. The smallest values
of $|\Vcb|$ and largest values of $\mt$ occur for large
 $\alphas$ and small $\mb$.  However, large values of $\alphas$
and small values of $\mb$  require large values for the
GUT scale top quark Yukawa, A.  For the range of parameters
studied in this paper $A<3$.  A perturbative one-loop calculation
becomes unreliable for larger $\alphas$ and smaller $\mb$.

Can we reduce the number of parameters at the GUT scale?
In table 1 we give some sample inputs at the GUT scale.

\bigskip
{\bf Table 1}
\vskip 0.25 cm
\begin{tabular}{|l|l|l|l|l|l|l|r|}    \hline \hline
$\tan{\beta}$ & A & B & C & D & E & F & $\phi$  \\ \hline \hline
  1.0 & 1.7 & .070 & .0002 & .01 & .0002 &  .00004  &   1.2
\\\hline
  10.0 &2.2  & .076 & .00014 & .070 & .001 &.0003   &   1.2
\\\hline
  60.0 & 1.3 & .05 &.001 & 1.46 & .014 &.0033   &  1.5  \\  \hline
\hline
\end{tabular}

\bigskip
The large mass hierarchy between similarly charged quarks in
different families requires  $A>>B>>C$ and $D>>E>>F$. However, since
the up and down quarks obtain  mass from the vacuum expectation
values of different Higgs doublets, there is a priori no
connection between A,B,C and D,E,F.
An extremely interesting question is whether the number of
predictions of the DHR scheme can be increased by arranging for relations
between the Yukawa couplings of the up and down sectors. A simple such relation
is A = D.  This occurs in $SO(10)$ models which have a single decouplet
generating the masses of the third generation provided that both light
$SU(2)$ doublets lie predominantly in this decouplet.

If attention is limited to the heaviest generation, the requirement that the
three Yukawa couplings are equal at the unification scale leads to a prediction
for the top mass \cite{ALS}: there are two parameters ($\lambda$ and $\tan
\beta$) and three observables. Unfortunately the prediction is very sensitive
to $\alpha_s$: $m_t$ increases from 110 GeV to near 180 GeV as $\alpha_s$
increases from .10 to .12. However, in the DHR framework the third generation
parameters cannot be treated in isolation, since they affect other observables,
such as $V_{cb}$. Setting A = D, ie making $\tan \beta$ very large, leads to
large values of $V_{cb}$, as can be seen from figure 2. Hence A = D is only
consistent with large $\alpha_s$. This considerably narrows our parameter
space: $\tan \beta = 60.6\pm 3 $, $m_t = 179\pm 4$ GeV, $\alpha_s \ge 0.
123$ and
$|V_{cb}| \ge .052$. Since the number of free parameters has been reduced by
one, $|V_{cb}|$ is now a prediction of the theory rather than an input.

Finally, we mention the impact of this framework on radiative
electroweak symmetry breaking in supersymmetric theories.     Recently,
Ananthanarayan, Lazarides and Shafi have demonstrated that
radiative electroweak symmetry  breaking can be accomplished
in the minimal supersymmetric model when
all three third generation Yukawa couplings are equal  at the
GUT scale\cite{ALS2}.
Since the DHR framework with $A=D$ tightly constrains $\mt$ and
$\tan{\beta}$ it leads to a very predictive heavy
sparticle mass spectrum.

\section{Conclusion}

In this paper we have given a general one-loop analysis of the DHR
framework.
The results of this work are shown in the Figures. Figures 2 and 4 show how the
predictions for $\tan \beta$ and $m_t$ depend on the input $V_{cb}$. In
particular, Figure 2 shows that for any inputs which DHR found led to an
acceptable  $\tan \beta$, there is also an additional large  $\tan \beta$
solution. Figure 3 shows that this large  $\tan \beta$ region always has a top
quark mass in the range 185$\pm$10 GeV, which is clearly much more restrictive
than in the region of lower $\tan \beta$.

The figures also provide an illustration of how the predictions for
$\tan \beta$ and $m_t$ depend on the inputs $m_c$, $m_b$ and especially
$\alpha_s$. It is because of the uncertainties in the values of these inputs
that $\tan \beta$ can vary so much. It is worth stressing that increasing
$\alpha_s(m_z)$ from .110 to .126 can reduce $m_t$ from 170 GeV to 126 GeV.

Restricting the DHR framework by imposing a relationship between the Yukawas of
the up and down sectors, namely setting A = D, produces a scheme where the
fourteen flavor parameters are predicted with just seven inputs. In this case
$V_{cb}$ is no longer an input, but is predicted by the theory. This very
predictive scheme can only be correct if $\alpha_s$ and $V_{cb}$ are both
large and if $m_t=179.\pm 4 $ GeV.

\indent
\vskip .5cm
We acknowledge useful conversations with Graham Ross.


\newpage
\begin{description}

\item[\it Figure 1:]
A plot of  the RGE mass enhancement factors $\eta_f$  versus  $\alphas(\mZ)$.
\item[\it Figure 2a:]
A plot of $\tan\beta$ versus $|\Vcb|$ for $\alphas(\mZ) = .110$,
On the solid (dashed) [dotted] curve, the  $\bar{MS}$ values of the
running quark masses
are $\mb(\mb) = 4.25 $ $(4.15)$ $[4.086]$ GeV and $\mc(\mc) = 1.27$ $(1.22)$
$[1.186]$ GeV.
\item[\it Figure 2b:]
A plot of $\tan\beta$ versus $|\Vcb|$ for  $\alphas(\mZ) = .118$.
The solid, dashed and dotted lines correspond to the same quark
masses as figure 2a.

\item[\it Figure 2c:]
A plot of $\tan\beta$ versus $|\Vcb|$ for  $\alphas(\mZ) = .126$.  The solid,
dashed and dotted lines correspond to the same quark masses as figure
2a.
\item[\it Figure 3a:]
A plot of the top quark's pole mass
$\mt$ versus $\tan{\beta}$ for  $\alphas(\mZ) = .110$;
inside the solid curve $,\mb(\mb) = 4.25 \pm .164$ GeV, $\mc(\mc)
 = 1.27 \pm .082
$ GeV, and $|\Vcb| \leq $.054.  With these restrictions, the top quark
mass is predicted to lie in the range  $170. <\mt <192.6$, while
$\tan{\beta}$ is restricted to $1.85< \tan{\beta} < 41$.  The
dotted curve near $\tan{\beta}=60.$ is the prediction when
$A=D$.

\item[\it Figure 3b:]
A plot of the top quark's
pole mass $\mt$ versus $\tan{\beta}$ for  $\alphas(\mZ) = .118$;
inside the dashed (solid) curve $,\mb(\mb) = 4.25 \pm .1 (\pm .164)$ GeV,
$\mc(\mc)= 1.27 \pm .05 (\pm .082)$
 GeV, and $|\Vcb| \leq .050 \,(\leq.054)$.  With these restrictions, the top
quark mass is
predicted to lie in the range $181.5 < \mt < 193.8$  $( 147 <\mt < 194.)$,
and $\tan{\beta}$ is restricted to  $2.6 < \tan{\beta} < 34.$
 $(1.15< \tan{\beta} < 56.)$ .

\item[\it Figure 3c:]
A plot of the top quark's
pole mass $\mt$ versus $\tan{\beta}$ for $\alphas(\mZ) = .126$;
inside the dashed (solid) curve $,\mb(\mb) = 4.25 \pm .1 (\pm .164)$ GeV,
$\mc(\mc) = 1.27 \pm .05 (\pm .082) $ GeV, and $|\Vcb| \leq .050 (\leq .054)
$.
With these restrictions, the top quark mass is predicted to lie in the range
 $155.4 < \mt < 195$ $( 126 <\mt < 195.5 )$, while $\tan{\beta}$
is restricted to  $1.3 < \tan{\beta} < 56.4$
 $( .83 < \tan{\beta} < 64.3 ) $. The dotted line gives the prediction
for the case $A=D$ with $\mc(\mc)$ = 1.188 GeV.

\item[\it Figure 4a:]
A plot of the pole mass $\mt$ versus $|\Vcb|$ for $\alphas(\mZ)= .110$.
The dotted, dashed and solid curves correspond to those
in figure 2.

\item[\it Figure 4b:]
A plot of the pole mass $\mt$ versus $|\Vcb|$ for $\alphas(\mZ) = .118$.
The dotted, dashed and solid curves correspond to those
in figure 2.
The additional dotted and dashed curve is for
$\mb(\mb) = 4.35 $ GeV, and $\mc(\mc) = 1.32$ GeV.  On each curve,
the arrows indicate the direction of increasing $\tan{\beta}$ and
monotonically decreasing, GUT scale top Yukawa, $A$.  The
circles indicate points  where $A = 2.0$.
\item[\it Figure 4c:]
A plot of the pole mass $\mt$ versus $|\Vcb|$ for $\alphas(\mZ) = .126$,
The dotted, dashed, and solid curves correspond to those
of figure 2.  The additional dotted and dashed curve is for
$\mb(\mb) = 4.35 $ GeV, and $\mc(\mc) = 1.32$ GeV.  On each curve,
the arrows indicate the direction of increasing $\tan{beta}$ and
monotonically decreasing, GUT scale top Yukawa, $A$.  The diamonds
(circles) indicate points where $A = 2.5 \, (2.0)$.
\end{description}

\end{document}